\documentclass[a4paper,11pt]{article}
\pdfoutput=1
\usepackage{jcappub} 
\usepackage[T1]{fontenc}
\usepackage{graphicx}
\usepackage{amsmath,amssymb}
\usepackage{hyperref}
\usepackage{url}
\usepackage{epstopdf}
\allowdisplaybreaks	
\numberwithin{equation}{section}
\usepackage{enumitem}
\usepackage{bm}
\usepackage[utf8]{inputenc}
\usepackage{color}
\usepackage{bm}
\usepackage{multirow}
\usepackage{hyperref}
\usepackage{subfigure}
\usepackage{array}
\usepackage[english]{babel}
\usepackage{tensor}
\usepackage{comment}
\usepackage[normalem]{ulem}
\usepackage[dvipsnames]{xcolor}

\hypersetup{
    colorlinks=true,
    linkcolor=black,
    citecolor=blue,
    urlcolor=black
}

\def\be{\begin{equation}}
\def\ee{\end{equation}}
\def\ba{\begin{eqnarray}}
\def\ea{\end{eqnarray}}
\def\nn{\nonumber}
\def\f{\frac}
\def\l{\left}
\def\r{\right}

\title{The role of the tachyonic instability in Horndeski gravity}

\author[a]{Noemi Frusciante}
\author[b]{Georgios Papadomanolakis}
\author[b]{Simone Peirone}
\author[b]{Alessandra Silvestri}

\affiliation[a]{Instituto de Astrof\'isica e Ci\^encias do Espa\c{c}o, 
Faculdade de Ci\^encias da Universidade de Lisboa,  Campo Grande, PT1749-016 Lisboa, Portugal}
\affiliation[b]{Institute Lorentz, Leiden University, PO Box 9506, Leiden 2300 RA, The Netherlands}

\emailAdd{nfrusciante@fc.ul.pt}
\emailAdd{papadomanolakis@lorentz.leidenuniv.nl}
\emailAdd{peirone@lorentz.leidenuniv.nl}
\emailAdd{silvestri@lorentz.leidenuniv.nl}

\abstract{The tachyonic instability is commonly associated  with   mass terms having a negative sign and 
evolving faster than the Hubble scale, leading to an unstable low-$k$ regime. In the cosmological exploration of modified gravity, it is seldom taken into account, with more focus given to the popular no-ghost and no-gradient conditions. The latter though are intrinsically high-$k$ statements. Here we combine all three conditions into a full set of requirements that we show to guarantee stability on the whole range of cosmological scales.
We then explore the impact of the different conditions on the parameter space of scalar-tensor gravity, with particular emphasis on the no-tachyon one.
We focus on Horndeski gravity and also consider separately the two subclasses of $f(R)$ and Generalized Brans Dicke theories. We identify several interesting features, for instance in the parameter space of designer $f(R)$ on a $w$CDM background, shedding light on previous findings. When looking at the phenomenological functions $\Sigma$ and $\mu$, associated to the weak lensing and clustering potential respectively,  we find that in the case of Generalized Brans Dicke the no-tachyon condition clearly cuts models with  $\mu\,,\,\Sigma>1$. This effect is less prevalent in the Horndeski case due to the larger amount of free functions in the theory. }

\begin{document}

\maketitle
\flushbottom

\section{Introduction}

The phenomenon of cosmic acceleration has given rise to an impressive bulk of theoretical models aimed at explaining the late time dynamics of the Universe. Out of these models, most are of the scalar-tensor form, with the additional scalar degree of freedom (dof) being associated either to a fluid-like component with negative pressure (DE) or a direct modification of General Relativity (MG). Their ultimate test will always be the comparison against data, but, at the theoretical level, there already exist conditions that a physically viable theory needs to satisfy~\cite{Bertacca:2007ux,Bertacca:2007cv,Gubitosi:2012hu,Bloomfield:2012ff,Gleyzes:2013ooa,Piazza:2013pua,Gleyzes:2014qga,Kase:2014cwa,Sbisa:2014pzo,Gergely:2014rna,Frusciante:2016xoj,Gumrukcuoglu:2016jbh,DeFelice:2017mwa,Lagos:2017hdr}. The ones considered most commonly are the no-ghost and no-gradient conditions, preventing the development of dofs with a negative kinetic term or a negative speed of propagation, respectively. Another, less explored, condition is the absence of tachyonic instabilities, related to the unboundedness of the Hamiltonian from below. 
In scalar-tensor theories, if one neglects matter fields, the tachyonic instability is related to the mass squared of the canonical scalar field. This instability manifests itself with a negative sign of the mass squared, yet when its rate of evolution is slower than the Hubble rate, one considers the theory to still be viable.
In $f(R)$ gravity~\cite{DeFelice:2010aj}, this corresponds to conditions on the mass of the \textit{scalaron}, and has been discussed in~\cite{Sawicki:2007tf}, in relation to the stability of the high curvature regime.

The presence of matter fields, as well as their potential mixing with scalar field, can not be neglected, especially in the context of the late time Universe. In this case, deriving the conditions to avoid tachyonic instabilities is quite involved and it requires finding the mass eigenvalues of the Hamiltonian of the canonical fields. 
Their expressions and the corresponding no-tachyon conditions have been worked out in~\cite{DeFelice:2016ucp} and they can be applied to a broad class of DE/MG models. Indeed, the authors used the model independent framework defined by the Effective Field Theory of dark energy (EFT of DE)~\cite{Gubitosi:2012hu,Bloomfield:2012ff} and the Sorkin-Schutz action~\cite{Schutz:1977df,Brown:1992kc} to model a dust fluid, e.g.~Cold Dark Matter (CDM). The final conditions where then obtained in the low-$k$ limit, where the tachyonic instability manifests.

In this paper, we implement the no-tachyon conditions into the stability module of the Einstein-Boltzmann solver \texttt{EFTCAMB}~\cite{Hu:2013twa,Raveri:2014cka,Hu:2014oga}, thus completing the already built-in no-ghost and no-gradient conditions, into a full set that, as we  show, guarantees stability on the whole range of linear scales. The already existing no-ghost and no-gradient, are in fact inherently high-$k$ statements, and as such cannot guarantee stability on the whole range of scales. The common practice in Einstein-Boltzmann solvers so far, was to include a set of mathematical conditions designed to eliminate models with exponential growth of the perturbations. The latter are ad-hoc conditions derived, under some simplifying assumptions, at the level of the equation of motion implemented in the numerical codes. Here we show that this will not be necessary anymore and one can rely on a rigorous, theoretically motivated, set of conditions to ensure stability both in the high-$k$ and low-$k$ regime. 

We then proceed to study the impact of the novel no-tachyon conditions on the parameter space of Horndeski~\cite{Horndeski:1974wa}, as well as of its subclasses $f(R)$ gravity and Generalized Brans Dicke theories, identifying several interesting features. Among other things, we confirm that in some cases the constraining power of the stability conditions can be significant and will certainly give an important contribution towards physically informed cosmological tests of gravity. 

The paper is organized as follows: In Sec.~\ref{Sec:Stability} we review the EFT of DE and the formulation of the stability conditions. In Sec.~\ref{Sec:Models} we present the class of models used in this study and their implementation in the EFT language. In Sec.~\ref{Sec:Methodology} we illustrate the methodology that allows us to build large ensembles of models. We also describe the parameter spaces that we use for studying the  impact of the different stability conditions. Finally, in Sec.~\ref{Sec:Results} we present and discuss the results and in Sec.~\ref{Sec:Conclusions} we conclude. 

\section{Stability conditions in the Effective Field Theory of dark energy}\label{Sec:Stability}

In this section we review the general conditions that a theory of gravity needs to satisfy in order to be free from instabilities, i.e.~no-ghost, no-gradient and no-tachyon conditions~\cite{Sbisa:2014pzo}. As discussed in the Introduction, we include matter fields in our derivation, focusing on CDM, which is the relevant one for late times.

We employ the EFT of DE framework, which is at the basis of the numerical code, \texttt{EFTCAMB}, that we use for our analysis. This framework offers a unified language for a broad class of DE/MG models with one additional scalar degree of freedom~\cite{Gubitosi:2012hu,Bloomfield:2012ff}. The corresponding action is constructed in the unitary gauge as a quadratic expansion in perturbations and their derivatives, around a flat Friedmann-Lema$\hat{\text{i}}$tre-Robertson-Walker (FLRW) background. The different operators that enter in the action are spatial-diffeomorphism curvature invariants. For the purpose of this paper we  restrict to Horndeski gravity, for which the corresponding EFT action reads:
\begin{align}\label{EFTaction}
&\mathcal{S}=\int d^4x\sqrt{-g}\l\{\frac{m_0^2}{2}(1+\Omega(t))R^{(4)}+\Lambda(t)-c(t)\delta g^{00}\r.\nn\\
&\l.+\frac{\bar{M}^2_3(t)}{2}\l[(\delta K)^2-\delta K^{\mu}_{\nu}\delta K^{\nu}_{\mu}-\f{1}{2}\delta g^{00}\delta R^{(3)}\r]\r.\l.+\frac{M^4_2(t)}{2}(\delta g^{00})^2-\frac{\bar{M}^3_1(t)}{2}\delta g^{00}\delta K+L_m[g_{\mu\nu},\chi_m]\r\},
\end{align}
where $m_0^2$ is the Planck mass, $g$ the determinant of the four dimensional metric $g_{\mu\nu}$, $\delta g^{00}$ the perturbation of the upper time-time component of the metric,  $R^{(4)}$ and $R^{(3)}$ are respectively the trace of the four dimensional and three dimensional Ricci scalar, $K_{\mu\nu}$ is the extrinsic curvature and $K$ its trace. $\Omega,c,\Lambda, M_i$ are free functions of time dubbed EFT functions. Finally, $L_m$ is the matter Lagrangian for all matter fields. In this work we strictly follow~\cite{DeFelice:2016ucp} and we adopt the Sorkin-Schutz matter Lagrangian, see~\cite{Schutz:1977df,Brown:1992kc}. 

After decomposing the action~(\ref{EFTaction}) into the actual perturbations of the metric and matter fields, and removing spurious dof, one obtains the following action for the propagating dofs in Fourier space~\cite{DeFelice:2016ucp}: 
 \begin{align}\label{actionshort}
\mathcal{S}^{(2)}=\int{} \f{d^3k}{(2\pi)^3}dta^3\l(\dot{\vec{\chi}}^t\textbf{A}
\dot{\vec{\chi}}-k^2\vec{\chi}^t\textbf{G}\vec{\chi}-
\dot{\vec{\chi}}^t\textbf{B}\vec{\chi}-\vec{\chi}^t\textbf{M}\vec{\chi}\r)\,,
\end{align}
where $\vec{\chi}^t\equiv (\zeta,\delta_d)$ with $\zeta$ the scalar degree of freedom and $\delta_d=\delta\rho/\bar{\rho}$ the density perturbations of the matter component. ${\bf A}$, ${\bf B}$, ${\bf G}$, ${\bf M}$ are $2\times2$ time and scale dependent matrices (see~\cite{DeFelice:2016ucp} for their expressions) and dots indicate time derivatives with respect to cosmic time.

The stability requirements can be now obtained from action~(\ref{actionshort})~\cite{DeFelice:2016ucp}. In the following we will list them, discussing their relevance and range of applicability: 
\begin{itemize}
\item \textit{no-ghost:} Requiring the absence of ghosts translates into $A_{ij}$ being positive definite. This must be done in the high-$k$ limit as a low-$k$ instability does not lead to a catastrophic vacuum collapse and is rather relatable to the Jeans instability, as discussed in~\cite{Gumrukcuoglu:2016jbh}. 


\item \textit{no-gradient:} In order to avoid diverging solutions at high-$k$ one needs to demand the speed of propagation to be positive, i.e. $c_s^2>0$. The speed of propagation can be obtained, after a diagonalization of the kinetic matrix $A_{ij}$, from the dispersion relations coming from action~\eqref{actionshort}. 


\item \textit{no-tachyon}: 
In the case of a single scalar canonical field, this amounts to demanding that either the mass term in the Lagrangian is positive or, in case it is negative, the rate of instability is slower than the Hubble rate. The latter case corresponds to the Jeans instability for the scalar dof.  When matter fields are involved, one needs to study the mass matrix of the Hamiltonian associated to the canonical fields as illustrated in~\cite{DeFelice:2016ucp}. For more details about the nature of the tachyon instability we refer the reader to~\cite{Joyce:2014kja}. Here, we will focus on the practical condition one has to impose in order to avoid such instability.

The Hamiltonian associated to the action~\eqref{actionshort} assumes the following general form for the canonical dofs~\cite{DeFelice:2016ucp}:
\begin{align}
\label{Hamiltonian}
\mathcal{H}=\frac{a^3}2\left[\dot{\Phi}_1^2 +\dot{\Phi}_2^2 +\mu_1(t,k)\,\Phi_1^2 +\mu_2(t,k)\,\Phi_2^2\right]\,, \nn\\
\end{align}
where $\Phi_i$ are the canonical fields and, for $k\rightarrow 0$, $\mu_i$ are the mass eigenvalues.\footnote{It is important to note that the canonical field is a result of a number of field redefinitions, hence is a mix of the scalar and matter dof.} 
As discussed in~\cite{DeFelice:2016ucp}, the above Hamiltonian exhibits a tachyonic instability when at low momenta a mass eigenvalue $\mu_i$ becomes negative and evolves rapidly, i.e. $|\mu_i| \gg H^2$. Thus, for a theory to be viable, it must hold that $|\mu_i|\lesssim H^2$. Formulated in this way one also includes theories with instabilities which evolve over scales much larger than the Hubble scale. This particular case is typical of clustering fluids, where the instability corresponds to the well known Jeans instability, which is vital for structure formation.
\end{itemize}

While we focused on the scalar dofs in the EFT action, one can repeat the same expansion for the tensorial part. Starting from the quadratic action for tensor perturbations, one can then work out the equivalent conditions to ensure that tensor modes are free from ghost and gradient instabilities (see e.g.~\cite{Gubitosi:2012hu,Bloomfield:2012ff,Gleyzes:2013ooa,Piazza:2013pua,Gleyzes:2014qga,Frusciante:2016xoj}). 

From the above discussion, we have five conditions (three for the scalar sector and two for the tensor one) which need to be imposed in order to guarantee the stability of a theory at any scale and time. 

The relevance of these conditions reflects also in the choice of the parameter space one has to sample when performing a fit to data. This was established in~\cite{Raveri:2014cka,Frusciante:2015maa,Peirone:2017lgi}, where it was shown that they might dominate over the constraining power of cosmological data. 

In Einstein-Boltzmann solvers the no-ghost and no-gradient conditions are commonly employed while the no-tachyon ones are typically not included. The former two conditions guarantee the stability at high-$k$, thus in order to guarantee stability on the whole $k$-spectrum, the codes usually employ ad-hoc conditions that eliminate models with exponentially growing modes at low-$k$. In the EFT framework, these additional requirements are typically worked out at the level of the dynamical equation for the perturbations of the scalar field. While in action~(\ref{EFTaction}) the scalar field is hidden inside the metric degrees of freedom, one can leave the unitary gauge by the St$\ddot{\text{u}}$ckelberg trick and make explicit the perturbations associated to the scalar field. This amounts to an infinitesimal time coordinate transformation $t \rightarrow t+\pi$, with the scalar degree of freedom being described by $\pi$ and obeying the following equations of motion:
\begin{align}
A \pi^{\prime\prime}+B\pi^\prime+C\pi+k^2 D\pi+H_0 E=0,
\end{align}
where $A,B,C,D,E$ are functions of time and k and their explicit expressions can be found in~\cite{Hu:2014oga}. $H_0$ is the present day value of the Hubble parameter and primes are derivatives with respect to conformal time.
 
The corresponding mathematical conditions are~\cite{Hu:2014oga}:
\begin{itemize}
\item if $B^2-4A(C+k^2D)>0$ then
\be
\label{MC1}
\f{-B\pm\sqrt{B^2-4A(C+k^2 D)}}{2A}<H_0\,,
\ee
\item if $B^2-4A(C+k^2D)<0$ then 
\be
\label{MC2}
\f{-B}{2A}<H_0\,.
\ee
\end{itemize}

The no-ghost and no-gradient conditions, as well as the mathematical ones, are implemented in the publicly available version of the Einstein-Boltzman solver \texttt{EFTCAMB}~\cite{Hu:2013twa,Hu:2014oga}, respectively under the name of \emph{physical} and \emph{mathematical} (math) conditions. Although it would be reasonable to add the no-tachyon conditions under the umbrella of physical conditions, in this paper we  stick to the original convention and retain the term physical for the pair of no-ghost and no-gradient. We  always refer separately to the no-tachyon one as the \emph{mass} condition.

In this work, we extend the stability module of \texttt{EFTCAMB} to include the mass conditions in terms of the mass eigenvalues $\mu_i$ and proceed to study the impact of the latter on the parameter space of different scalar-tensor theories within Horndeski gravity. Our first goal is to show that the physical plus mass conditions form a complete set of physically motivated, rigorously derived requirements that do guarantee stability on the whole range of linear cosmological scales. We also compare the mass and math conditions in terms of performance, showing that the latter can be safely disregarded in favor of the former. Finally we study the effects of the different conditions on the parameter space of the different theories, identifying some noteworthy features. 

\section{Models}\label{Sec:Models}
We consider several classes of scalar-tensor models of gravity:
\begin{itemize}[]
\item f(R)~\cite{DeFelice:2010aj}: specifically designer $f(R)$~\cite{Song:2006ej,Pogosian:2007sw} with a $w$CDM background. For any value of the equation of state, $w_0$, the different models reproducing the corresponding expansion history can be labeled by the present value of the Compton wavelength of the scalaron, namely $B_0=B(z=0)$, where
\begin{equation}
\label{ComptonWave}
B(z)=\frac{f_{RR}}{1+f_R}\frac{H \dot{R}}{\dot{H}}\,.
\end{equation} 
Hence we  have a two-dimensional model parameter space, i.e. $\{w_0,B_0\}$. These models can be fully mapped in the EFT language, and the only corresponding non-zero EFT functions are: $\Omega=f_R$ and $\Lambda=\frac{m_0^2}{2}\left(f-Rf_R\right)$. 
\item Generalized Brans Dicke (GBD): non-minimally coupled scalar-tensor theories with a canonical kinetic term. f(R) gravity is a sub-class of these theories, with a fixed coupling to matter. When one allows the coupling to vary, a representative class is that of Jordan-Brans-Dicke (JBD) models~\cite{Brans:1961sx}. These models correspond, in the EFT language, to non-zero $\{\Omega, \Lambda, c\}$. In this work we  adopt the so-called `pure EFT' approach, where we  simply explore several different choices for $\{\Omega, \Lambda, c\}$ as functions of time, thus creating a large ensemble of GBD models. For more details on this approach we refer the reader to~\cite{Raveri:2017qvt,Peirone:2017ywi}.
\item Horndeski (Hor): the full class of second order scalar-tensor theories as identified by Horndeski~\cite{Horndeski:1974wa}. Within the EFT formalism, we can explore them by turning on the full set of EFT functions in action~(\ref{EFTaction}), i.e. $\{\Omega, \Lambda, c, M_2^4,\bar{M}_1^3,\bar{M}_3^2\}$.
We  also consider separately the subset of Horndeski for which the speed of sound of tensor is equal to that of light, $c_t^2=1$. Most of the modifications happen in the scalar sector of the theory, hence we  refer to this class as $H_S$. This specific class has become of great interest after the detection of the gravitational wave GW170817 and its electromagnetic counter part GRB170817A~\cite{TheLIGOScientific:2017qsa,Monitor:2017mdv,Coulter:2017wya}, which has set tight constraints on the speed of propagation of tensor modes. We can create large ensembles of these models in the pure EFT approach, by turning on the following EFT functions: $\{\Omega, \Lambda, c, M_2^4,\bar{M}_1^3\}$.

\end{itemize}

\section{Methodology}\label{Sec:Methodology}

We aim at studying in detail the way different sets of stability conditions affect the parameter space of the models under consideration. We always impose  the set of physical stability conditions, i.e. no-ghost and no-gradient,  as a baseline; on top, we separately switch on the checks for either the math~(\ref{MC1} or~\ref{MC2}) or the mass condition.  For one class of  models, namely $f(R)$, we consider an additional condition, as desrcibed in the following. 

f(R)-gravity is among the models for which the stability conditions have been extensively investigated~\cite{Amendola:2006we,Song:2006ej,Pogosian:2007sw,Hu:2007nk,Starobinsky:2007hu}. Besides the usual no-ghost and no-gradient conditions, an additional important requirement has been identified in the literature by demanding the high curvature regime to be stable against small perturbations. This translates into requiring a positive mass squared for the scalaron, in the limit of $|Rf_{RR}|\ll1$ and $f_R \rightarrow 0$
\be \label{massfR}
m^2_{f_R}\approx \f{1+f_R}{3f_{RR}} \approx \f{1}{3f_{RR}}\,,
\ee
which leads to the well known $f_{RR}>0$ constraint, often dubbed as the tachyon condition for $f(R)$. Since such condition has been obtained under specific hypothesis, it does not coincide analytically with the general no-tachyon conditions discussed in Sec.~\ref{Sec:Stability}. In the analysis of $f(R)$, we include this latter condition as one of the case studies, to compare with the mass and math ones.

To study the viable parameter space of all the models under the different sets of stability conditions, we use the numerical framework adopted in~\cite{Raveri:2017qvt,Peirone:2017ywi}. It consists of a Monte Carlo (MC) code which samples the space of the EFT functions, building a statistically significant ensemble of viable models.
To compute wether a sampled model is stable (and thus accepted by the sampler) or not, we interface the MC code with the publicly available Einstein-Boltzmann solver \texttt{EFTCAMB}~\cite{Hu:2013twa,Raveri:2014cka}. Starting from the background solution, the code undergoes a built-in check for the stability of the model.

In \texttt{EFTCAMB}, $f(R)$ is implemented via the so-called `mapping' mode, i.e. as a specific model after being mapped to the EFT language~\cite{Hu:2014oga}. Currently both the Hu-Sawicki model~\cite{Hu:2016zrh} and designer $f(R)$ models are available. For our study we focus on the latter one, choosing a $w$CDM background. 
For the remaining three classes of models we adopt the `pure EFT' approach, where we explore many different choices for the time-dependence of the corresponding EFT functions.
Specifically, following~\cite{Raveri:2017qvt,Peirone:2017ywi,Espejo:2018hxa}, we parametrize the relevant EFT functions using a Pad\'e expansion:
\begin{align}
f(a) = \frac{\sum_{n=1}^{N} \alpha_{n} \left( a-a_0\right)^{n-1}}{1+\sum_{m=1}^{M} \beta_m \left( a-a_0\right)^m}\ ,\label{def_pade}
\end{align}
where the truncation orders are given by $N$ and $M$. The coefficients $\alpha_{n}$ and $\beta_m$ are sampled with uniform prior in the range $[-1,1]$ and we verified that the results are not sensitive to the prior range. Furthermore, the convergence of the results is reached at $N+M=9$, meaning that each EFT function has $9$ free parameters. We consider, with equal weight, expansions around $a_0=0$ and $a_0=1$ to represent thawing and freezing models, respectively. For further details about the sampling procedure we refer the reader to~\cite{Raveri:2017qvt,Peirone:2017ywi}. 

In this pure EFT approach any choice of $\Lambda$ and $\Omega$ produces a different background expansion history, which can be solved for using the Friedmann equation, as explained in~\cite{Raveri:2017qvt}. The remaining background EFT function, $c$, can be determined in terms of $\Lambda, \Omega$ and $H(a)$, and does not need to be sampled independently. 

Following this procedure, we build large numerical samples of viable models for the GBD, $H_S$ and Hor classes reaching a samples size of $\sim 10^4$ accepted models.

For $f(R)$ gravity we study the impact of stability conditions on the $\{w_0,B_0\}$ parameter space, comparing also to previous results~\cite{Raveri:2014cka}. For the remaining classes of models the dimension of the parameter space is very high (e.g.  GBD has $27$ additional parameters) and, furthermore, the individual parameters in the Pad\'e expansion do not have any direct physical meaning. For such reasons we look at their predictions for the phenomenology of Large Scale Structure, studying the cuts of different stability criteria on the, physical, space of the phenomenological functions $(\mu, \Sigma)$ defined in the usual way~\cite{Bean:2010zq}:
\ba
 k^2 \Psi &=& -4\pi G \mu (a,k) a^2 \rho \Delta \ ,
\label{poisson_mg} \\
 k^2(\Phi +\Psi) &=& -8\pi G\, \Sigma(a,k)\, a^2 \rho \Delta \ ,
\label{weyl_mg}
\ea
where $\rho$ is the background matter density, $\Delta=\delta +3aHv/k$ is the comoving density contrast, and $\Psi$ and $\Phi$ are the scalar perturbations, respectively, to the time-time and spatial components of the metric in conformal Newtonian gauge. From their definition, $\mu=\Sigma=1$ in $\Lambda$CDM, but in general they are functions of time and scale. The function $\mu$ directly affects the clustering and the peculiar motion of galaxies, hence it is well constrained by galaxy clustering and redshift space distortion measurements~\cite{Song:2010fg,Simpson:2012ra,Asaba:2013mxj}. On the other hand, $\Sigma$ affects the geodesics of light and is directly measured by Weak Lensing, Cosmic Microwave Background and galaxy number counts experiments~\cite{Moessner:1997qs,Hojjati:2011xd,Asaba:2013mxj}. 

\begin{figure}
\begin{center}
\includegraphics[width=1.\textwidth]{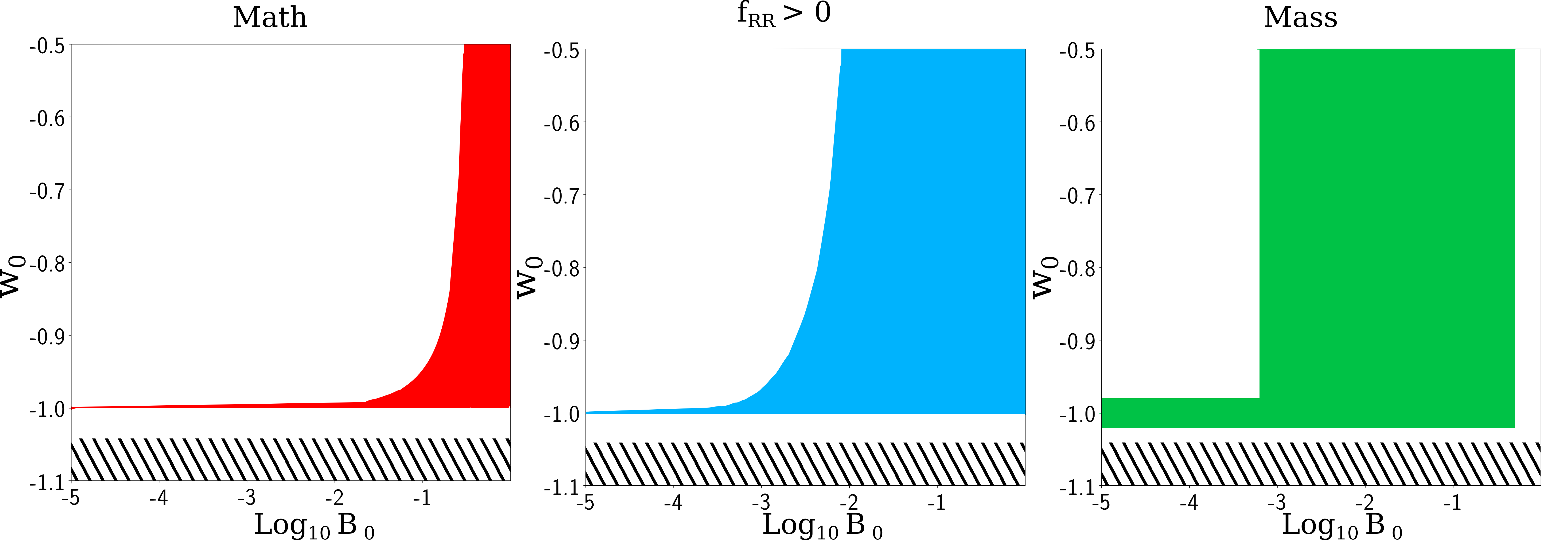}
\caption{The viable $(B_0,w_0)$ parameter space of designer $f(R)$ gravity on a $w$CDM background. We plot the regions allowed by different stability conditions. The physical conditions are imposed as a baseline in all cases and cut the lower region highlighted by black lines. On top of them, we apply separately the math (left panel), $f_{RR}>0$ (central panel) and the mass (right panel) conditions. The corresponding viable regions are indicated with solid colors, respectively in red for math, blue for $f_{RR}>0$ and green for mass. 
}\label{fig:B0_w0}
\end{center}
\end{figure}

The \texttt{EFTCAMB} software allows us, in principle, to evolve the full dynamics of linear perturbations and extract the exact form of $\Sigma$ and $\mu$ for each model in our ensembles. This however would be highly time-consuming, hence we opt for the Quasi Static (QS) analytical expressions of these functions, worked out from the modified Einstein equations after reducing them to an algebraic set (in Fourier space) by neglecting time derivatives of the scalar degrees of freedom~\cite{Bloomfield:2012ff,Pogosian:2016pwr}. In~\cite{Peirone:2017ywi} the authors have compared the QS and exact $(\Sigma,\mu)$ for the same ensembles of models as those considered in this paper, finding that the agreement is excellent for all scales below the typical Compton wavelength of the sampled model. 

In order to visualize the effect of different stability cuts, we show the predictions of a given ensemble of models in the $(\Sigma, \mu)$ plane for a given value of scale and redshift.
We set our output scale at $k = 0.01h$ Mpc$^{-1}$, which has been shown to be safely inside the Compton scale for all models considered in the present analysis~\cite{Peirone:2017ywi}. For this scale, we show the results at a given value of the scale factor that we choose to be $a=0.9$, corresponding to a redshift $z \approx 0.1$. This choice is more realistic than $a=1$ in terms of measurements from upcoming surveys.

\section{Results}\label{Sec:Results}

We now proceed to discuss the outcome of our analysis, focusing on the cuts in the different parameter spaces considered. When studying the GBD and Horndeski classes,  through the large ensemble of models generated via the Monte Carlo sampling, we  also analyze the acceptance rates of models when the different conditions are turned on. 
\begin{figure}[t]
\begin{center}
\includegraphics[width=.4\textwidth]{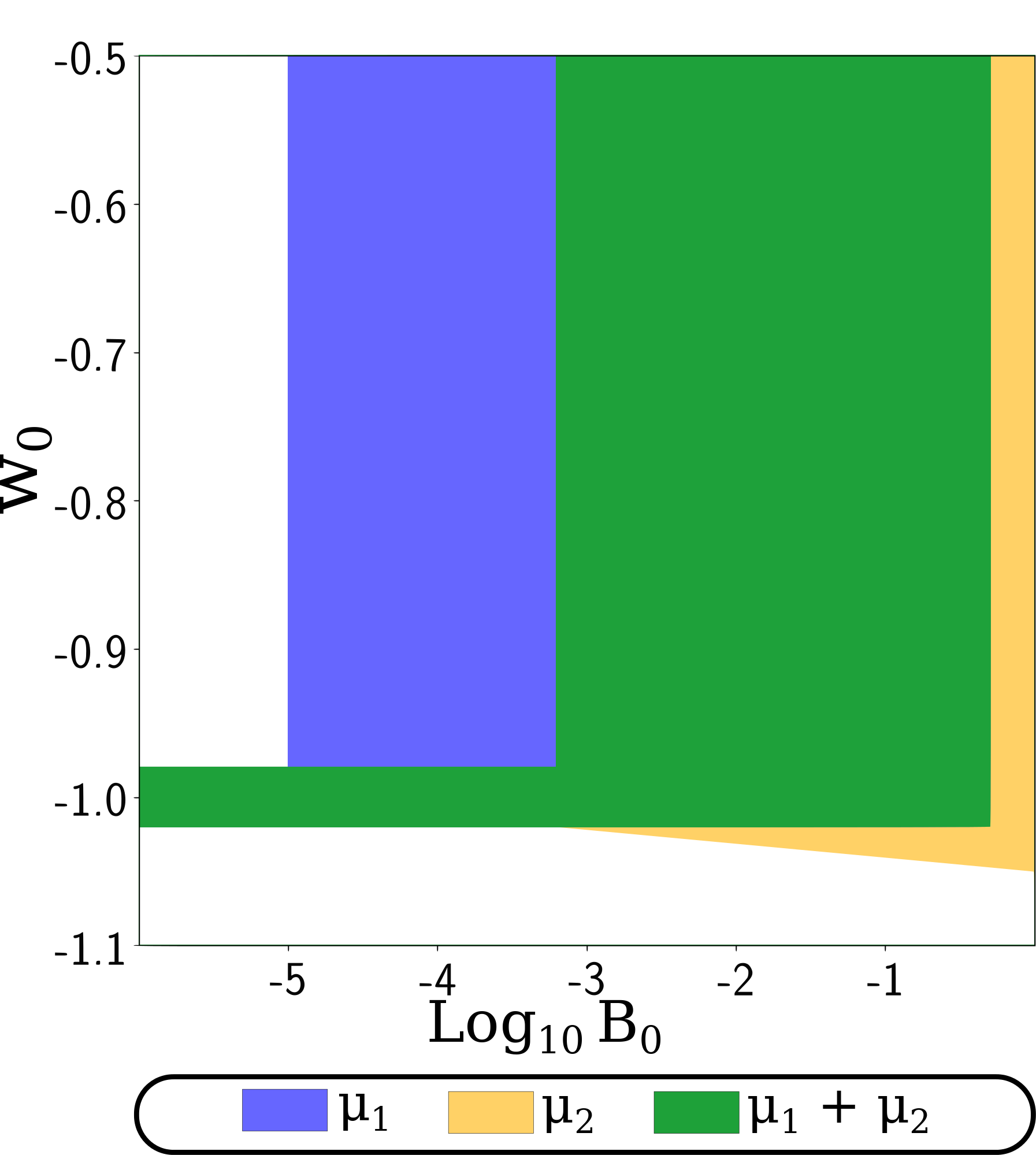}
\caption{Allowed parameter space for designer $f(R)$ gravity with $w$CDM background, when applying the physical conditions jointly with the mass conditions considered separately, i.e. the $\mu_1$ (blue) or $\mu_2$ (yellow) stability cuts. In green we show the combined viable region.
\label{fig:fR_m12}}
\end{center}
\end{figure}

\subsection{Designer $f(R)$ on $w$CDM background}\label{Sec:fR}

We studied designer $f(R)$ on a $w$CDM background, considering flat priors on $w_0 \in [-1.1, -0.5]$ and $Log_{10} \, B_0 \in [-5,0]$, while fixing the cosmological parameters to Planck 2015 $\Lambda$CDM values~\cite{Ade:2015xua}. We  also tested that our results do not change when assuming the values from Planck 2018~\cite{Aghanim:2018eyx}, as they are mostly compatible with the 2015 release.

In Fig.~\ref{fig:B0_w0} we show the impact of the math, $f_{RR}>0$ and mass conditions on the $(B_0,w_0)$ space. As it was already known, the physical conditions do not constrain $B_0$ while they clearly constrain the equation of state to $w_0> -1.04$. Going beyond the baseline, it is evident that the math condition constrains the parameter space most severely, pushing $w_0$ towards $w_0=-1$ for values of $B_0\lesssim 10^{-2}$. On the other hand, $f_{RR}>0$ and mass condition have a very similar, and less severe, impact.  We have studied the cosmology of a number of models excluded by the math conditions but allowed by the $f_{RR}$ and mass conditions. They all exhibited stable behaviors, hence we infer the math conditions for $f(R)$ are too stringent.  

The viable parameter space of the same designer $f(R)$ was studied in~\cite{Raveri:2014cka} under the no-ghost, no-gradient and the $f_{RR}>0$ requirements. Their findings are in line with our results.  Let us now look at Fig.~\ref{fig:fR_m12}, where we consider separately the constraints coming from the individual mass eigenvalues. One thing that can be noticed, is that both mass conditions are  important, cutting regions of the parameter space that are partially complementary. The combined effect is similar to that  of the $f_{RR}$ condition, and this is a direct result of the mixing of the scalar and matter dof, which plays an important role in the determination of the full no-tachyon conditions.   

\begin{figure*}[t!]
\begin{center}
\includegraphics[width=.8\textwidth]{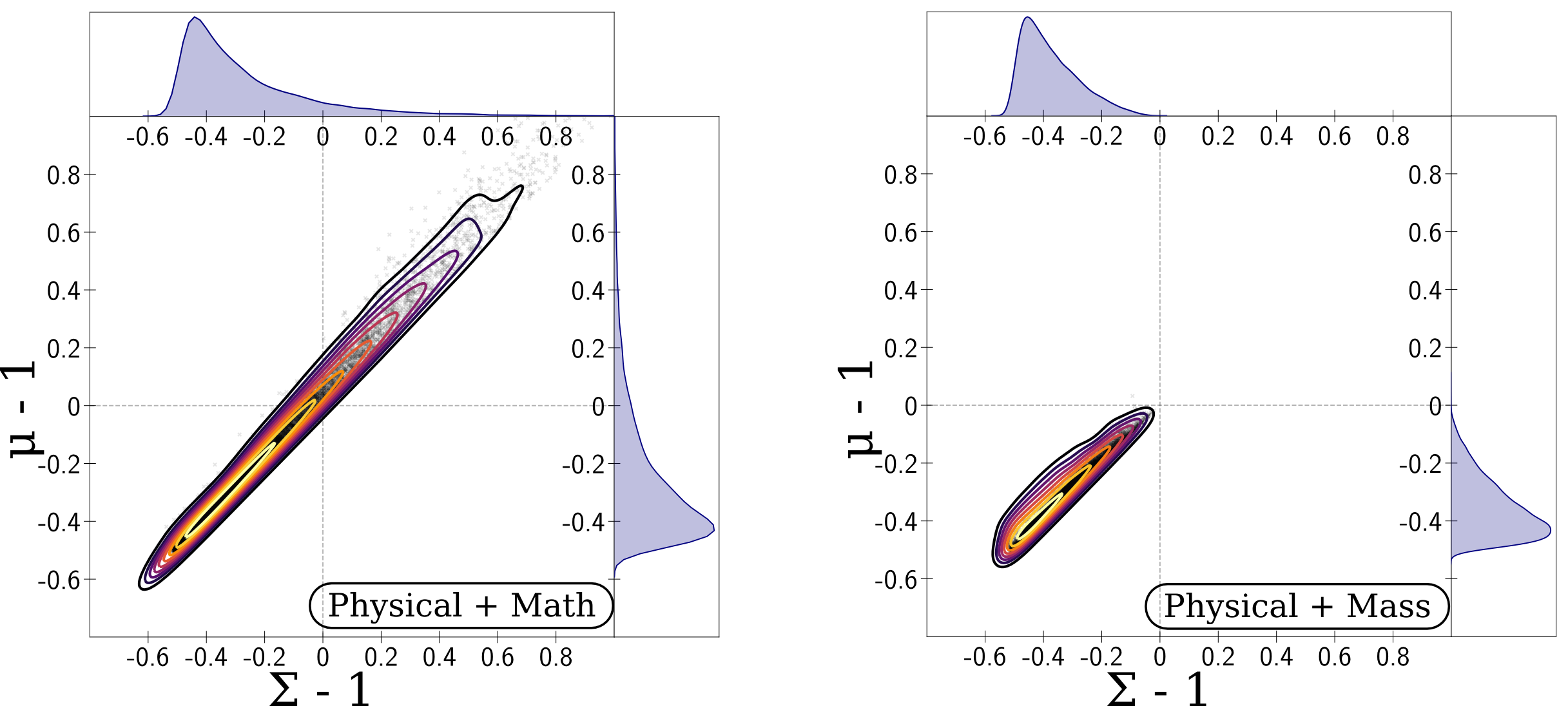}
\caption{Marginalized 2D and 1D distributions for the phenomenological functions $\Sigma -1$ and $\mu -1$ for the GBD models, computed at $a=0.9$ and $k = 0.01 h$Mpc$^{-1}$. In the two panels we show the results of the different stability cuts: on the \textit{left} physical+math and on the \textit{right} physical+mass.
Black shaded points represent the values computed for the single sampled models, while the contour lines cut the 2D distribution at $10$\%, $20$\%, \dots, $90$\% of the total sample. These results are computed within the QSA. 
\label{fig:GBD_Sigmamu}}
\end{center}
\end{figure*}

\begin{figure*}[t!]
\begin{center}
\includegraphics[width=.8\textwidth]{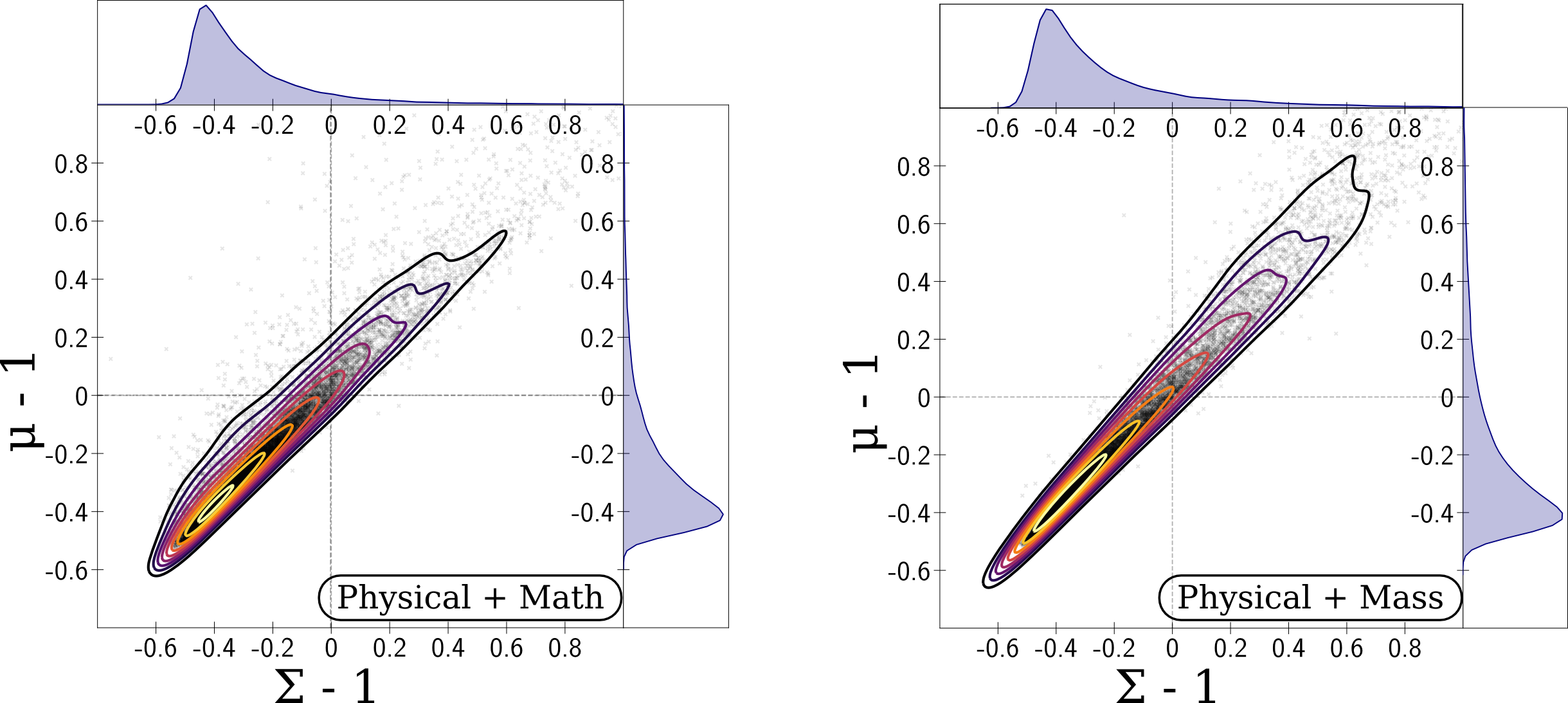}
\includegraphics[width=.8\textwidth]{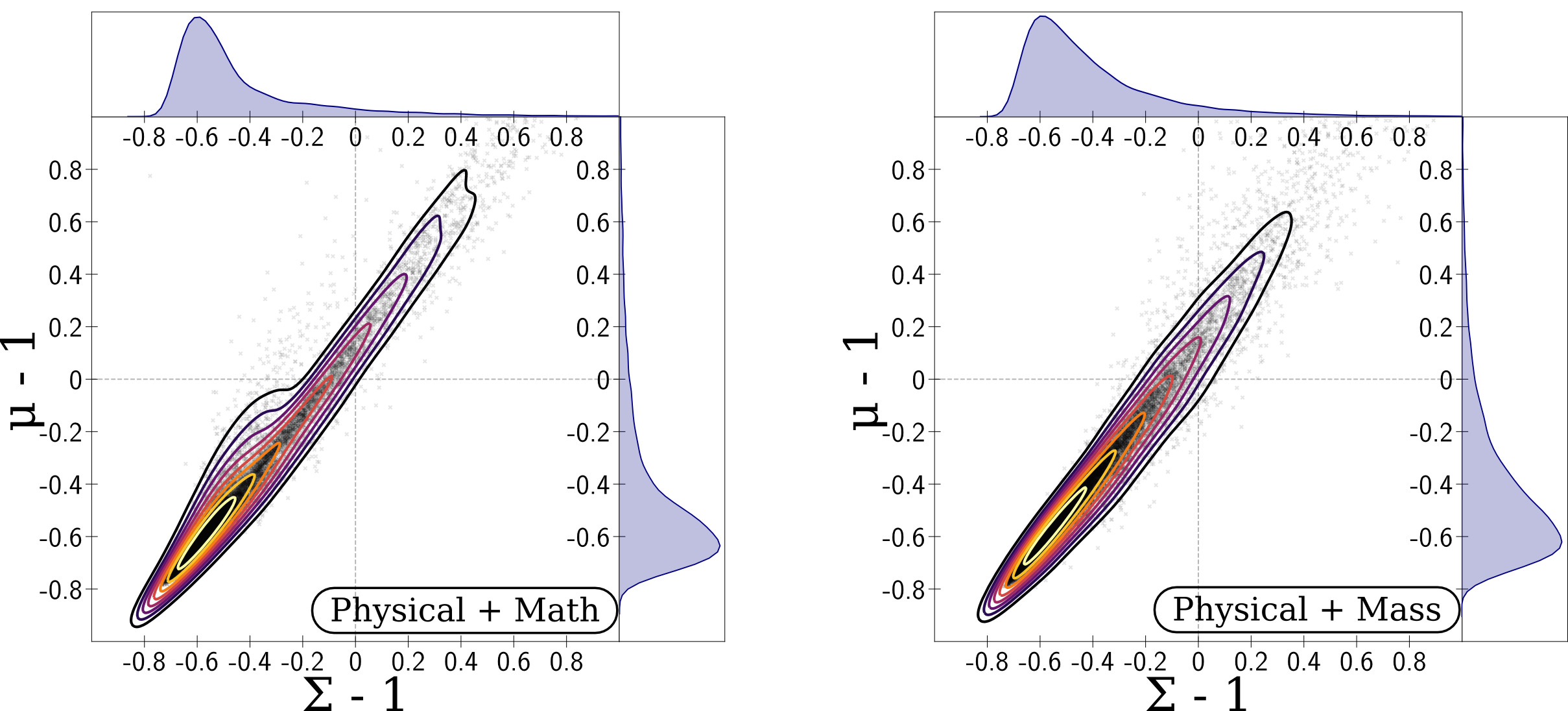}
\caption{Marginalized 2D and 1D distributions for the phenomenological functions $\Sigma -1$ and $\mu -1$ for $H_S$ (top panels) and Hor (bottom panels), computed at $a=0.9$ and $k = 0.01 h$Mpc$^{-1}$. In the two panels we show the results of the different stability cuts: on the \textit{left} physical+math and on the \textit{right} physical+mass.
Black shaded points represent the values computed for the single sampled models, while the contour lines cut the two dimensional distribution at $10$\%, $20$\%, \dots, $90$\% of the total sample. These results are computed within the QSA.
\label{fig:Horndeski_Sigmamu} }
\end{center}
\end{figure*}

\subsection{Horndeski}\label{Sec:Horndeski}

The results for Generalized Brans Dicke models (GBD), Horndeski with $c_t^2=1$ ($H_S$) and full Horndeski (Hor) are presented as marginalized 2D and 1D distributions for the phenomenological functions $\Sigma-1$ and $\mu-1$, at $a=0.9$ and $k = 0.01 h$Mpc$^{-1}$. In all cases the sampling was done till $10^4$ models were accepted by the set of stability conditions under consideration. 

In Fig.~\ref{fig:GBD_Sigmamu} we show the marginalized 2D and 1D distributions for $(\Sigma,\mu)$ for the ensemble of GBD models, when applying the math and mass stability cuts, on top of the usual physical baseline. It becomes instantly clear that the mass condition has a stronger impact on the plane than the math conditions. While the math conditions generally allow models satisfying the relation $(\mu-1)(\Sigma-1)>0$, the mass conditions break the degeneracy and allow purely models with $(\mu-1),(\Sigma-1)<0$.\footnote{This boundary when imposed by the math condition is rather sharp, while imposed by the mass condition can be violated by a statistically negligible number of models ($0.01$\% of the total sample).} In terms of EFT functions this result translates into the mass conditions requiring
$\Omega>0$, since in the QSA $\Sigma\sim \frac{1}{1+\Omega}$. The difference between mass and math conditions is quite relevant when analyzed in the $(\Sigma,\mu)$ plane, yet in terms of acceptance rates, the two conditions do not differ much, as shown in Table~\ref{acceptance_rates}. In retrospect, one notices that in fact the bulk of viable models were in the $(\mu-1),(\Sigma-1)<0$ quadrant already for the math conditions case. Still, it is noteworthy that the mass conditions in GBD models forbid more distinctively the first quadrant. 

In Fig.~\ref{fig:Horndeski_Sigmamu} the same combinations of stability conditions are studied in the phenomenological plane of $H_S$ and Hor models. In this case, the difference between the mass and math conditions is not so evident in terms of allowed regions in the $(\Sigma, \mu)$ plane. Nevertheless, we can notice that the math conditions allow the ensembles of models to have a tail in the second quadrant ($\mu >1$ and $\Sigma <1$), while this tail is drastically cut by the requirement of mass stability. In fact the models lying in the second quadrant are reduced from $\sim 1$\% in the former case to $\sim 0.1$\% in the latter.

As we discussed in Sec.~\ref{Sec:Methodology}, we adopt a Monte Carlo sampling technique to create ensembles of models that obey different sets of stability conditions. It is quite informative to compare the acceptance rates for different stability conditions and we present their percentage values in Table~\ref{acceptance_rates}. It can be immediately noticed that the baseline of physical conditions has a quite strong impact on Horndeski models, a result previously discussed in~\cite{Peirone:2017ywi}. Focusing on the mass and math conditions a general trend emerges, the mass conditions are more stringent than the math conditions. For GBD and $H_S$, this effect is stronger than for Hor where the impact of the two conditions is fairly similar. Finally, let us notice that, while in the $(\Sigma,\mu)$ plane of $H_S$ and Hor there were no striking differences between mass and math conditions, looking at the acceptance rates we do see some tangible differences. This can be easily understood in terms of the large number of free EFT functions that describe these models. In other words, mass conditions do cut more models than math conditions, yet when we look at the phenomenology, i.e. at the $(\Sigma,\mu)$ plane, there is high degeneracy among the models and we can not associate specific regions to these cuts. This is in stark contrast to  GBD where the additional cut contributed by mass has a specific direction in the $(\Sigma,\mu)$ plane due to the fact that mainly one EFT function, $\Omega$, is being constrained.
\begin{table}
\begin{center}
\begin{tabular}{ |l|c|c|c|}
	\hline
	&GBD (\%)& $H_S$ (\%)& Hor (\%) \\[1mm]
	\hline 
	physical & $18$ & $8.3$ & $ 1.2$ \\[.5mm]
	physical + math & $15$ & $1.9$ & $0.3$ \\[.5mm]
	physical + mass & $13$ & $1.1$ &$0.2$\\[1mm]
	\hline
\end{tabular}
\end{center}
\caption{Acceptance rates for the GBD, $H_S$ and Hor classes of models as subjected to the different sets of stability requirements: physical, physical + math and physical + mass.\label{acceptance_rates}}
\end{table}

\section{Conclusions}\label{Sec:Conclusions}

We have investigated the impact of the condition for the avoidance of tachyon instabilities in scalar-tensor theories, throughout referred to as \emph{mass} condition. Such condition is crucial to guarantee the stability of theories on the whole range of linear scales, complementing the no-ghost and no-gradient conditions. The latter are the ones most commonly implemented in Einstein-Boltzmann solvers, but being intrinsically high-$k$ conditions, they can not guarantee stability on all scales. So far this shortcoming was addressed with a set of \emph{mathematical} conditions built on the additional scalar field equation, in such a way to filter out models with exponentially growing solutions that would have escaped the no-ghost and no-gradient check. 

A complete derivation of the mass condition in Horndeski, and more general modified gravity models, was carried out in~\cite{DeFelice:2016ucp}. We have used the theoretical results of that paper as the basis for our analysis, implementing the corresponding conditions in the stability module of \texttt{EFTCAMB}. We then have carried out an extensive study of the impact of the new conditions on the parameter space of Horndeski gravity, in particular comparing the corresponding cuts to those previously contributed by the mathematical conditions, as well as the improvement brought upon the incomplete set of no-ghost and no-gradient. 

Overall, we show that, as expected, the mass condition provides the missing constraining power at low-$k$, that up till now was achieved by the mathematical conditions. Combined with the no-ghost and no-gradient, they form a complete set of conditions that guarantee the stability of any theory over all cosmological scales. Additionally, while the mathematical conditions depended on a number of simplifying assumptions about the nature of the dynamical equation for the extra degree of freedom, the mass conditions are more rigorously defined from a physical point of view. They are obtained from the stability analysis of the full action and do not rely on any such simplifying assumptions. 

In our analysis, we considered separately different scalar-tensor theories, as implemented in \texttt{EFTCAMB}, i.e. $f(R)$ gravity, Generalized Brans Dicke (GBD) models, the full Horndeski action, as well as the subset of it that does not alter the speed of tensor modes. In all cases, we used the combination of no-ghost and no-gradient as the baseline and, on top of that, compared the performance of mathematical versus mass conditions. As mentioned above, the mass condition proves to be a very reliable substitute of the mathematical condition in all cases. On top of this, there are some features peculiar to specific models that are worth summarizing. 
One of the most evident results of this work is that, in GBD models, the mass condition removes more efficiently models away from the $ \Sigma\,,\,\mu>1$ region, clearly cutting the tail of models that were allowed by the mathematical conditions. This indicates that GBD models with either $\mu$ and/or $\Sigma$ bigger than one would develop a tachyon instability. While this feature is interesting and clearly stands out in the $(\Sigma,\mu)$ plane of Fig.~\ref{fig:GBD_Sigmamu}, it is important to notice, from the acceptance rates in Table \ref{acceptance_rates}, that the tail is a small fraction of the whole ensemble of models, that instead tends to live in the $\Sigma<1\,,\,\mu<1$ quadrant. Interestingly the latter would be severely constrained if one imposes consistency with local tests of gravity, as shown in~\cite{Peirone:2017ywi}.

It is also worth stressing that in the case of $f(R)$ gravity, we compared the mass condition not only with the mathematical one, but also with the popular $f_{\rm R R}>0$ condition, which is based on the stability of the theory in the high-curvature regime. As shown in Fig.~\ref{fig:B0_w0}, we found that the ad-hoc mathematical condition is too stringent in the $f(R)$ case, while the $f_{\rm RR}>0$ and mass ones contribute an almost equivalent, and more generous, cut to the parameter space.

For the full Horndeski class, as well as the sub-class obeying $c_t^2=1$ at all times, we do not report significant differences between the impact of mathematical and mass conditions. But we highlight that the mass condition completes the no-ghost and no-gradient into a reliable set of conditions that guarantees stability on all linear scales, while being physically informed. 

As we have shown, the combination of no-ghost, no-gradient and no-tachyon forms a theoretically rigorous and practically important set of conditions that guarantees stability on all linear cosmological scales. Finally, let us notice that, while in this work we focused on Horndeski gravity, the mass conditions derived in~\cite{DeFelice:2016ucp} and implemented in the stability module, can cover beyond Horndeski models as well. 

 \begin{acknowledgments}
We are grateful to Levon Pogosian, Marco Raveri  for useful discussions. The research of NF is supported by Funda\c{c}\~{a}o para a  Ci\^{e}ncia e a Tecnologia (FCT) through national funds  (UID/FIS/04434/2013), by FEDER through COMPETE2020  (POCI-01-0145-FEDER-007672) and by FCT project ``DarkRipple -- Spacetime ripples in the dark gravitational Universe" with ref.~number PTDC/FIS-OUT/29048/2017.  GP, SP and AS acknowledge support from the NWO and the Dutch Ministry of Education, Culture and Science (OCW), and also from the D-ITP consortium, a program of the NWO that is funded by the OCW. The authors acknowledge the COST Action (CANTATA/CA15117), supported by COST (European Cooperation in Science and Technology).
\end{acknowledgments}

\bibliographystyle{aipnum4-1}
\bibliography{biblio_G}

\end{document}